\documentstyle[pra,aps]{revtex}

\begin{document}
\title{Collapses and revivals in the interference between 
two Bose-Einstein
condensates formed in small atomic samples}
\author{E. M. Wright\\
Optical Sciences Center\\
University of Arizona\\
Tucson, AZ 85721, USA\\
[0.1cm] T. Wong, M.J. Collett, S.M. Tan and D. F. Walls\\
Department of Physics\\
University of Auckland\\
Private Bag 92019\\
Auckland, New Zealand}
\date{\today}
\maketitle

\begin{abstract}
We investigate the quantum interference between two 
Bose-Einstein condensates formed in small atomic
samples composed of a few thousand atoms both by
imposing Bose broken gauge symmetry from the outset 
and also using an explicit model of atomic detection.
In the former case we show that the macroscopic wave
function collapses and revives in time, and we calculate
the characteristic times for current experiments. Collapses 
and revivals are also predicted in the interference
between two Bose-Einstein condensates which are initially
in Fock states, a relative phase between the condensates
being established via atomic detections corresponding to 
uncertainty in the number difference between them.

\vspace{0.2cm} \noindent PACS numbers: 03.75.Fi, 05.30.Jp, 
32.80.Pj, 74.20.D
\end{abstract}

\newpage

\section{Introduction}

The spectacular observations of Bose-Einstein 
condensation (BEC) in atomic vapors reported last year
\cite{AndEnsMat95,DavMewAnd95} have opened up new
avenues of research into the physical
properties and nature of Bose-condensed systems.
Recent experimental developments include reports
of a new trap capable of holding larger number of
atoms and measurements of the condensate fraction
and mean-field energy \cite{MewAndDru96a},
non-direct observation of the development of the
condensate \cite{MewAndDru96b}, measurements of
the collective oscillations of the condensate
\cite{JinEnsMat96}-\cite{JinMatEns96},
and an output coupler for an atomic
Bose-Einstein condensate \cite{MewAndKur96}.
The measurements of the collective excitations have been found
to be in excellent agreement with the theoretical
predictions from mean-field theory for condensate
fractions near unity \cite{Fet96}-\cite{Str96}.
Such detailed studies of the collective excitations
opens the door to investigating superfluid effects in
atomic BECs, and in particular the general relation between
BEC and superfluidity (see, for example, the article
by Huang in Ref. \cite{GriSnoStr95}).

In this paper we investigate the intereference of two
BECs formed in small atomic samples composed of
$10^3-10^6$ atoms, typical of current experiments.
Our motivation for this theoretical investigation
is that we previously showed that in small atomic
samples the macroscopic wave function exhibits
collapses and revivals, the first collapse
occuring on a few seconds time scale
\cite{WriWalGar96}.  In that work the notion of
Bose broken symmetry \cite{Gri93} was invoked
and the quantum state of the condensate was
taken as a wave packet composed of states of
fixed number of particles, so that, according to
the uncertainty relationship between particle number
and phase, the
condensate phase can become well defined.
The collapse of the macroscopic
wave function arises from the fact that,
due to many-body interactions, the chemical
potential is different for each particle
number present in the wave packet, and the
revivals are a direct consequence of the
discreteness of the particle number
\cite{WriWalGar96}.  Based on this theory
one would naively expect that if
two BECs are interfered
the visibility of the interference pattern
should also exhibit the collapses and revivals,
hence providing a direct experimental signature
of the effect.  However, before jumping to
conclusions it is important to note that
the notion of Bose broken symmetry is
only strictly applicable in the thermodynamic
limit, where the system size and the number of
particles tend to infinity with the density
held constant, so that another approach should be
utilized to verify the predicted collapse and
revivals.  Such an approach, freed from the
thermodynamic limit and Bose broken symmetry
arguments, has emerged in the last year
\cite{JavYoo96}-\cite{Mol96}.  Javanainen and Yoo
\cite{JavYoo96} first showed that an interference
pattern between two condensates, and hence a relative
phase between the two condensates, could arise from
an explicit model of atomic detections: as atomic
detections are performed an uncertainty in the
number of atoms in each condensate is built up,
since we don't know which condensate any atom
is removed from.  The uncertainty relationship
between the relative atom number of the
condendates and the relative phase then allows
a well defined relative phase to emerge.

The remainder of this paper is organized as follows:
Section II gives the basic theory for BEC in a
trapped gas of atoms as described by the
Gross-Pitaevskii equation and within the Thomas-Fermi
approximation.  The interference between two condensates
using the notion of Bose broken symmetry is described
in Sec. III, and the collapse and revival times for
the macroscopic wave function are evaluated for
current experimental conditions.  In Sec. IV we
approach the interference of two condensates using a
quantum anharmonic oscillator approximation to the full
field-theoretical model of Sec. II, and
the atomic detection scheme of Ref. \cite{JavYoo96}.
Here we verify that the collapse and revivals are still
present without explicitly invoking Bose broken symmetry.
In Sec. V we briefly discuss the quantum anharmonic
oscillator model for condensates described by
coherent states, ie. assuming Bose broken symmetry,
and demonstrate that the details of the collapses
and revivals depends on the quantum state of the
condensate.  Finally, Sec. VI gives our summary and
conclusions.

\section{Basic Theory}

In this Section we introduce the second quantized 
Hamiltonian for BEC in a
trapped gas of atoms, and then discuss the Gross-Pitaevskii 
equation for the
macroscopic wave function and the Thomas-Fermi approximation. 
Although these
topics are amply covered elsewhere we include them for 
completeness and to
clarify our notation.

\subsection{Second quantized Hamiltonian}

Our starting point is the second quantized Hamiltonian for 
a system of
bosonic atoms confined in a trap potential \cite{Hua87} 
\begin{equation}
\hat{H}(t)=\int d{\bf r}\left[ \frac{\hbar ^{2}}{2m}\nabla 
\hat{\psi}%
^{\dagger }\cdot \nabla \hat{\psi}+V({\bf r})\psi ^{\dagger 
}\hat{\psi}+%
\frac{U_{0}}{2}\hat{\psi}^{\dagger }\hat{\psi}^{\dagger }
\hat{\psi}\hat{\psi%
}\right] ,  \label{QHam}
\end{equation}
where $\hat{\psi}({\bf r},t)$ and $\hat{\psi}^{\dagger }
({\bf r},t)$ are the
Heisenberg picture field operators which annihilate and 
create atoms at
position ${\bf r}$, and obey the equal-time commutation 
relation $[\hat{\psi}%
({\bf r},t),\hat{\psi}^{\dagger }({\bf r}^{\prime },t)]
=\delta ({\bf r}-{\bf %
r}^{\prime })$ appropriate to bosons, the (single-particle) 
trap potential
is taken of the form \cite{EdwBur95,RupHolBur95,BayPet96} 
\begin{equation}
V({\bf r})=\frac{1}{2}m(\omega _{\perp }^{0})^{2}(r_{\perp }
^{2}+\lambda
^{2}z^{2}),  \label{trapot}
\end{equation}
with ${\bf r}=({\bf r}_{\perp },z),{\bf r}_{\perp }$
being the transverse position
coordinate assuming cylindrical symmetry of the trap 
potential in the
transverse plane, and $z$ the longitudinal coordinate, $m$ 
is the atomic
mass, $\omega _{\perp }^{0}$ the transverse angular 
frequency of the trap, $%
\lambda =\omega _{z}^{0}/\omega _{\perp }^{0}$ is the ratio 
of the
longitudinal to transverse frequencies, and $U_{0}=4\pi 
\hbar ^{2}a/m$
measures the strength of the two-body interaction, $a$ 
being the s-wave
scattering length. Here we consider a repulsive interaction 
so that $a>0$.

\subsection{Gross-Pitaevskii equation}

Here we consider the standard treatment of the condensate 
in a Bose gas \cite
{LifPit89}. At zero temperature, and for a weakly 
interacting gas, the
particles may be assumed to be predominantly in the 
condensate.  The assumption of zero temperature is
not too restrictive since atomic condensates can
be prepared with condensate fractions close to
unity \cite{JinEnsMat96,MewAndDru96c}.  The
Schr\"odinger equation for the state
vector of the system is 
\begin{equation}
i\hbar \frac{\partial}{\partial t}|\Psi(t)\rangle =\hat 
H(0)|\Psi(t)\rangle ,
\label{SchroEq}
\end{equation}
and in the time-dependent Hartree approximation
\cite{Dir30,KerKoo76,YooNeg77}
the state vector for a system of $N$ particles
is written as 
\begin{equation}
|\Psi(t)\rangle=(N!)^{-1/2}\left[\int d{\bf r} \psi_{N}
({\bf r},t)\hat\psi%
^\dagger ({\bf r},0)\right]^N|0\rangle ,
\end{equation}
where $\psi_{N}({\bf r},t)$ is the normalized 
single-particle wave function,
and $|0\rangle$ the vacuum state. Then proceeding in the 
usual manner \cite
{KerKoo76,YooNeg77} from the Schr\"odinger equation 
(\ref{SchroEq}) we
obtain the self-consistent nonlinear Schr\"odinger equation 
or
Gross-Pitaevskii equation \cite{LifPit89} generalized to 
include the magneto-optical harmonic trap 
\cite{EdwBur95,RupHolBur95,BayPet96}
\begin{equation}
i\hbar \frac{\partial \psi _{N}}{\partial t}=\left[ 
-\frac{\hbar ^{2}}{2m}%
\nabla ^{2}+\frac{1}{2}m(\omega _{\perp }^{0})^{2}(r_{\perp 
}^{2}+\lambda
^{2}z^{2})+NU_{0}|\psi _{N}|^{2}\right] \psi _{N},  
\label{GPEq}
\end{equation}
We are interested in the ground state solution of this 
equation for which we
set $\psi _{N}({\bf r},t)=\exp (-i\mu _{N}t/\hbar )\phi _{N}
({\bf r})$,
giving the stationary Gross-Pitaevskii equation 
\begin{equation}
\mu _{N}\phi _{N}=\left[ -\frac{\hbar ^{2}}{2m}\nabla ^{2} 
+\frac{1}{2}%
m(\omega _{\perp }^{0})^{2}(r_{\perp }^{2}+\lambda ^{2}z^{2}
 )+NU_{0}|\phi
_{N}|^{2}\right]\phi _{N},  \label{GroundEq}
\end{equation}
where the macroscopic wave function $\phi_N({\bf r})$
 for the $N$-particle system is normalized to unity.
The ground-state energy of the system
of $N$ atoms is derived from the Ginzburg-Pitaevskii-Gross 
energy functional \cite{GPGfunc}
\begin{equation}
{\cal E}_N=N\int d{\bf r} \left [\frac{\hbar^2}{2m} 
|\nabla\phi_N|^2+V({\bf r%
})|\phi_N|^2 +\frac{NU_0}{2}|\phi_N|^4\right ] ,  
\label{Efunc}
\end{equation}
and the parameter $\mu_{N}$ is given by $\mu_{N}=d{\cal E}
_{N}/dN$. By a
slight extension of the usual terminology $\mu_{N}$ will be 
called the
chemical potential of the N-particle condensate. The 
stationary
Gross-Pitaevskii equation follows from the variational 
principle $\delta (%
{\cal E}_N-N\mu_N\int d{\bf r}|\phi_N|^2)=0$, and using 
this we find the further exact result 
\begin{equation}
\mu_N^{\prime}=\frac{d\mu_N}{dN} =U_0\int d{\bf r}
|\phi_N|^4 +\frac{NU_0}{2}
\int d{\bf r}\frac{\partial}{\partial N}|\phi_N|^4  ,
\end{equation}
which can be rearranged to yield
\begin{equation}
U_0\int d{\bf r}|\phi_N|^4
=\left (1+\frac{N}{2}\frac{\partial}{\partial N}\right )^{-1}
\mu_N^{\prime}  ,
\label{overlap1}
\end{equation}
which we shall use below.

\subsection{Thomas-Fermi approximation}

Edwards and Burnett \cite{EdwBur95}, and Ruprecht et. al. 
\cite{RupHolBur95}
have solved Eq. (\ref{GroundEq}) numerically to obtain both 
the condensate
wave functions and the chemical potentials $\mu_N$ as 
functions of $N$. Here
we follow Baym and Pethick \cite{BayPet96} and use the 
Thomas-Fermi
approximation for the macroscopic wave function, in which 
the kinetic energy
term in the stationary Gross-Pitaevskii equation is 
neglected. This yields
the approximate macroscopic wave function 
\begin{equation}
\phi_N({\bf r}_\perp,z)
=(NU_0)^{-1/2} \sqrt{\mu_N-\frac{1}{2} 
m(\omega_\perp^0)^2
(r_\perp^2+\lambda^2 z^2)} ,
\end{equation}
when the argument of the square root is greater than or 
equal to zero, zero
otherwise. Requiring that the macroscopic wave function be 
normalized to
unity yields the expression for the chemical potential 
\begin{equation}
\mu_N=\frac{\hbar\omega_\perp^0}{2} \left(\frac{15\lambda 
Na}{a_\perp}%
\right)^{2/5} ,
\label{ChemPot}
\end{equation}
where $a_\perp=(\hbar/m\omega_\perp^0)^{1/2}$ is the 
characteristic
transverse length scale of the linear potential. The 
Thomas-Fermi
approximation is valid if the coherence length $r_{coh} 
=(8\pi an)^{-1/2}$ 
\cite{LifPit89}, $n$ being the mean density, is less than 
the characteristic
size $r_N$ of the atomic cloud. Following Baym and Pethick 
\cite{BayPet96}
we find that 
\begin{equation}
\zeta=\frac{r_{coh}}{r_N}= \left(\frac{15\lambda Na}
{a_\perp} \right)^{-2/5}
,
\end{equation}
which should be much less than unity for the Thomas-Fermi 
approximation to
apply. In the first four columns of Table I we show the 
values of the
parameters for some current experiments with a 
positive scattering
length, and the fifth column shows the corresponding values 
of $\zeta$, and
we see that the Thomas-Fermi theory should be valid in all 
cases.

The explicit form of the chemical potential in Eq.
(\ref{ChemPot}) can be used to evaluate the overlap
integral in Eq. (\ref{overlap1}) giving
\begin{equation}
U_0\int d{\bf r}|\phi_N|^4
=\frac{10}{7}\mu_N^{\prime}  ,
\label{overlap}
\end{equation}
a result we shall use later.

\section{Quantum interference with Bose broken symmetry}

In this Section we investigate quantum interference between 
two BECs using
the notion of Bose broken gauge symmetry \cite{Gri93}.
According to this notion the
macroscopic wave function for each condensate attains a 
fixed but arbitrary
phase as a result of the spontaneous breaking of U(1) gauge 
symmetry during
the condensation process. The U(1) gauge symmetry is 
associated with
particle number conservation, so that in the condensed 
system the particle
number is not fixed. Then, as a result of the 
number-phase uncertainty
relationship $\Delta N\Delta \eta \approx 1$, the phase 
$\eta $ need not be
indeterminate, and interference between two BECs is 
possible. Here we employ
a description of the condensate state vector as a 
wavepacket of states of
fixed particle number N in the condensate to apply the 
notion of Bose broken
gauge symmetry. In contrast, in Section \ref{section3} we 
will show atomic
detections establish a coherent wavepacket without the need 
to build in Bose broken symmetry from the outset.

\subsection{Wavepacket description}

The definition of the macroscopic wave function given in the 
last Section is
rigorous in the thermodynamic limit \cite{LifPit89}. To 
investigate the case
of a small condensate, say a few thousand atoms, we employ 
a wavepacket
description. In particular, the wavepacket description is 
intended to
reflect the quantum coherence of the condensate for 
measurement times short
compared to relaxation times \cite{BarBurVac96,HohMar65}. 
In contrast, for measurements performed over times
long compared to relaxation times the quantum coherence
of the condensate is destroyed and the macroscopic
wave function vanishes.

Here we employ a wavepacket composed of states of fixed 
number of particles $ N$ in the condensate,
with expansion coefficients
$a_{N}=|a_{N}|e^{i\zeta_{N}}$, hence retaining quantum
coherence. The present description of
BEC, due to Barnett et. al. \cite{BarBurVac96}, is 
therefore different from
the conventional $\eta $-ensemble which employs a 
wave-packet of states
corresponding to different total number of particles, 
condensate plus
non-condensate, and is generally only applicable in the 
thermodynamic limit 
\cite{HohMar65}. A coherent state with $a_{N}=\bar{N}^{N/2}
e^{iN\eta }e^{-\bar{N}/2}/\sqrt{N!}$
suggests itself, but the states associated with BEC do
not generally possess such complete phase-coherence 
\cite{And84}.
Nevertheless, a pure state description of the condensate 
may be rendered
plausible as follows: Below the Einstein condensation 
temperature the
many-body ground state of the system becomes 
macroscopically occupied,
yielding a condensate which should be considered an open 
quantum system in
contact via many-body interactions with the environment or 
reservoir
composed of the non-condensate atoms. It is known from the 
work of Zurek et.
al. \cite{ZurHabPaz93} and Gallis \cite{Gal96}, that for a 
system, in their
case a harmonic oscillator, in interaction with an 
environment certain pure
states show considerable stability against loss of quantum 
coherence, and
that in the weak coupling limit the pure states of maximal 
stability are the
coherent states. Thus, the quantum coherence of the 
condensate may be
reasonably represented by a pure state, though perhaps not 
precisely a
coherent state since weak coupling may not hold. This 
argument does not
depend on the size of the system, except that the 
non-condensate atoms may
be viewed as a reservoir, and the conclusions therefore 
apply even for small
condensates far removed from the thermodynamic limit.

We further assume that the particle number distribution 
$|a_{N}|^{2}$ is
sharply peaked with variance $\Delta N$ around a mean 
particle number $\bar N
$. As a concrete example we take a Poissonian distribution 
for which $\Delta
N={\bar{N}}^{1/2}$, which is reasonable since the number 
distribution $%
|a_{N}|^{2}$ of the condensate should be approximately that 
of a coherent
state, though the phases $\zeta _{N}$ may not be so 
correlated. Under these
conditions we may expand the chemical potential $\mu_{N}$ 
around the mean
number $\bar{N}$ in a Taylor series 
\begin{equation}
\mu_{N}=\mu _{\bar{N}}+(N-\bar{N})\mu _{\bar{N}}^{\prime } 
+\frac{1}{2}(N-%
\bar{N})^{2}\mu _{\bar{N}}^{\prime \prime }+\ldots ,  
\label{First_pert}
\end{equation}
where 
\begin{equation}
\mu _{\bar{N}}^{\prime }=\frac{3}{5}\hbar \omega _{\perp }
^{0}\left( \frac{%
\lambda a}{a_{\perp }}\right) ^{2/5}\frac{1}{\bar{N} ^{3/5}}
 .
\end{equation}
If we compute the ratio of the third term to the second term
in this expansion for $N=\bar N+\Delta N$, then we find
that the magnitude of this ratio varies as
$\bar N^{-1/2}$.  Thus, even for a low particle number
$\bar N=10^3$ the second term
in the expansion is the dominant correction.  In the remainder
of this paper we shall therefore only retain the first two terms
in the above expansion.  

\subsection{Condensate state vector}

We are now equipped to construct the wavepacket for the 
condensate state
vector, which we write as 
\begin{equation}
|\Psi (t)\rangle =\sum_{N}a_{N}e^{-iN\mu_{N}t/\hbar} 
(N!)^{-1/2}\left(\hat{a}^{\dagger}\right)^{N}|0\rangle,
\label{Wave_Func}
\end{equation}
where the exponential contains the term $N\mu_{N}$ since 
there are $N$
particles each of chemical potential $\mu_{N}$, the 
operator $\hat{a}^{\dagger }=\int d^{3}{\bf r}
\phi_{\bar{N}}({\bf r} )\hat{\psi}^{\dagger}({\bf r},0)$ 
creates particles with distribution $\phi 
_{\bar{N}}({\bf r})$ 
\cite{Hua87}, with $[{\hat{a}},{\hat{a}}^{\dagger }]=1$, 
and $|0\rangle $ is
the vacuum state. The approximations employed in Eq. 
(\ref{Wave_Func}) are
tantamount to the Hartree approximation. In general the 
Schr\"odinger field
annihilation operator can be written as a mode expansion 
over
single-particle states as 
\begin{eqnarray}
\hat{\psi}({\bf r},0)&=&\sum_{\alpha} \hat a_{\alpha}
\varphi_{\alpha}({\bf r}%
)  \nonumber \\
&=&\hat{a}\phi _{\bar{N}}({\bf r}) +\tilde{\psi}({\bf r},0) 
,
\end{eqnarray}
where $\{\varphi_{\alpha}({\bf r})\}$ are a complete 
orthonormal basis set.
Here in the last line we have chosen $\varphi_{0}({\bf r}
)=\phi_{\bar N}(%
{\bf r})$ as one member of the complete orthonormal set, 
and identified $%
\hat a=\hat a_0$. Then by construction the first term in 
the mode expansion
acts only on the condensate state vector, whereas the 
second term $\tilde{%
\psi}({\bf r},0)$ accounts for the non-condensate atoms.

\subsection{Collapses and revivals}

The state vector (\ref{Wave_Func}) may be used to calculate 
quantum
expectation values of the condensate. Of interest to us 
here are the
one-particle reduced density matrix representing the 
condensate 
\begin{eqnarray}
\rho _{1}({\bf r},{\bf r}^{\prime },t) &=&\langle \Psi 
(t)|\hat{b}^{\dagger }%
\hat{b}|\Psi (t)\rangle \phi _{\bar{N}}^{*}({\bf r})\phi 
_{\bar{N}}({\bf r}%
^{\prime })  \nonumber \\
&=&\bar{N}\phi _{\bar{N}}^{*}({\bf r})\phi _{\bar{N}}({\bf 
r}^{\prime }),
\label{rho_wave}
\end{eqnarray}
and the macroscopic wave function 
\begin{eqnarray}
\langle \hat{\psi}({\bf r},t)\rangle &=&\langle \Psi 
(t)|\hat{b}|\Psi
(t)\rangle \phi _{\bar{N}}({\bf r})  \nonumber \\
&=&\bar{N}^{1/2}\phi _{\bar{N}}({\bf r})e^{-i\mu t/\hbar } 
{\cal F}_{\bar{N}%
}(t),  \label{phi_wave}
\end{eqnarray}
where 
\begin{equation}
{\cal F}_{\bar{N}}(t)=\sum_{N}\sqrt{\frac{N}{\bar{N}}} 
a_{N-1}^{*}
a_{N}[\cos (2\mu _{\bar{N}}^{\prime }(N-\bar{N})t/\hbar ) 
-i\sin (2\mu _{%
\bar{N}}^{\prime }(N-\bar{N})t/\hbar )],  \label{Foft}
\end{equation}
and $\mu =\mu _{\bar{N}}+{\bar{N}}\mu _{\bar{N}}^{\prime }$
is the net chemical potential of the condensate.
Using Eq. (\ref{ChemPot}) for the chemical potential
$\mu_N$
in the Thomas-Fermi approximation we obtain
$\mu=\frac{7}{5}\mu_{\bar N}={\cal E}_{\bar N}/\bar N$,
and the net chemical potential $\mu$
is equal to the mean
energy per particle \cite{BayPet96}.
In obtaining these 
results we have
used the first two terms in the expansion of the chemical 
potential (\ref
{First_pert}). The one-particle reduced density matrix 
(\ref{rho_wave}) is
of the classic factorized form representing the
off-diagonal-long-range order (ODLRO)
associated with BEC \cite{PenOns56,Yan62}.
This conclusion holds for any mean particle number, as long 
as the mean
field approximations employed are valid. Here the ODLRO 
extends over
separations $|{\bf r}-{\bf r}^{\prime}|\approx r_{\bar{N}}$,
the spatial
scale of the condensate. In the experiments the 
one-particle reduced density
matrix is not measured but rather the momentum space 
(velocity) distribution
is obtained by releasing the atoms from the trap, letting 
them fall under
gravity, and imaging them. The momentum spread will then be 
$\Delta K\approx
2\pi /r_{\bar{N}}$, so that the ODLRO is transferred to a 
sharp spike in the
imaged atomic distribution \cite{AndEnsMat95,DavMewAnd95}.

Turning now to the macroscopic wave function
(\ref{phi_wave}), we see that
the factor ${\cal F}_{\bar{N}}(t)$ takes the form of a 
weighted sum of
trigonometric functions with different frequencies. Such 
sums are well known
from the Jaynes-Cummings model of quantum optics, which 
describes the
interaction between a single-mode radiation field and a 
two-level atom, and
give rise to the phenomenon of collapses and revivals 
\cite{EbeNarSan80},
and the same is expected here. Collapse and revivals also 
appear in the
relative phase between two superfluids or superconductors 
\cite{Sol94}.
Directly from the form of Eq. (\ref{Foft}) we see that 
$|{\cal F}_{\bar{N}}(t)|$ is periodic in time
with period
\begin{equation}
T_{\bar{N}}=\pi\hbar /\mu _{\bar{N}}^{\prime },
\label{RevPer}
\end{equation}
and the revivals occur with this period. The 
revivals result
from the fact that the sum in Eq. (\ref{Foft}) is over the 
discrete particle
number, so that they are a direct result of the granularity 
of matter. The
collapses depend on the choice of the number distribution 
$|a_{N}|^{2}$. For
our purposes we only need the variance $\Delta N$ in 
particle number and the
assumption that the phases are correlated enough that 
${\cal F}_{\bar{N}}(0)$
does not vanish exactly. Then ${\cal F}_{\bar{N}}(t)$ is 
periodic and has a
maximum in magnitude at $t=t_{max}$. At this time the net 
phases for each $N$
are such that they add most constructively in the sum in 
Eq. (\ref{Foft}).
As time increases these phase relations will initially be 
lost thus
producing collapse of the magnitude of ${\cal F}_{\bar{N}} 
(t)\rightarrow 0$%
, until the system revives at $t=t_{max}+T_{\bar{N}}$. The 
collapse time $%
t_{coll}$ may be estimated by looking at the spread of 
frequencies present
in the wavepacket for particle numbers between $N={\bar{N}} 
\pm \Delta N/2$,
which yields $\Delta \Omega =2\mu _{\bar{N}}^{\prime } 
\Delta N/\hbar $, and 
$t_{coll}\approx 2\pi /\Delta \Omega $. Gathering our 
results together for
the collapse and revival times we have 
\begin{equation}
T_{\bar{N}}\approx \frac{5}{\omega_{\perp}^0}\left( 
\frac{a_{\perp }}{\lambda a}%
\right) ^{2/5}\bar{N}^{3/5},\qquad t_{coll}\approx 
\frac{T_{\bar{N}}}{\Delta
N}.  \label{times}
\end{equation}
The collapse phenomenon actually occurs under far more 
general conditions
than reflected by the approximations used here, the 
essential ingredient
being dispersion of the chemical potential $\mu _{N}$ over 
the particle
number variance $\Delta N$. Landau damping of plasma 
oscillations in an
electron plasma is another example of decay of a coherent 
state. In
contrast, the exactly periodic revivals arise from the 
linear dependence of $%
\mu _{N}$ on $(N-\bar{N})$ employed in Eq. (\ref{First_pert}
 ). If
higher-order corrections are retained in the expansion 
(\ref{First_pert})
the revivals are no longer perfectly periodic, and diminish 
with increasing
time. We also note that according to Eq. (\ref{rho_wave}) 
the
single-particle reduced density matrix is insensitive to 
the collapse and
revivals of the macroscopic wave function, which is then 
also the case for
the atomic BEC experiments \cite{AndEnsMat95,DavMewAnd95}.

\subsection{Thermodynamic limit}

The thermodynamic limit of these results must be taken with 
care. In
particular, in order to maintain a constant density as the 
mean particle
number is increased it is necessary to concomitantly 
increase the linear
trap size as $r_t\propto{\bar N}^{1/6}$. Then $T_{\bar N} 
\propto {\bar N}$
and $t_{coll}\propto{\bar N}^{1/2}$, so that collapse and 
revivals become irrelevant for $\bar N\rightarrow\infty$.
In addition, we find that ${\cal F}_{\bar N} 
(0)\rightarrow
e^{i\eta}$ by the following argument: The approximate 
uncertainty relation $%
\Delta N\Delta\eta\approx 1$ holds for the number and phase 
fluctuations of
the condensate. Then, as $\Delta N={\bar N}^{1/2} 
\rightarrow\infty$ we have 
$\Delta\eta\rightarrow 0$, which is the case for a coherent 
state with
phases $\zeta_N=N\eta$. Thus, in the thermodynamic limit 
the quantum state
of the condensate approaches a coherent state for which we 
find ${\cal F}_{%
\bar N}(0)=e^{i\eta}$. We then have $\langle\hat\psi({\bf r}
,t)\rangle
=e^{i\eta}{\bar N}^{1/2}\phi_{\bar N}({\bf r} )e^{-i\mu 
t/\hbar}$, and this
is precisely the limit in which the macroscopic wave 
function acts as an
order parameter \cite{LifPit89}.

\subsection{Interference between two condensates}

We now consider the case that the atomic BEC is interfered 
with a second
large condensate whose macroscopic wave function we write as 
$C\exp [i({\bf k}%
\cdot {\bf r}-\mu t/\hbar +\chi )]$. Here we have assumed 
that the second
condensate is spatially large and write it as a plane-wave 
with wavevector $%
{\bf k}$, and has the same chemical potential as the first 
for simplicity.
We further assume that the second condensate is composed of 
a large number
of atoms so that collapses and revivals are not an issue, 
the constant $C$
is real, and $\chi $ is the arbitrary but fixed phase of 
the condensate.
Then, when the two condensates are made to interfere, say 
by dropping them
on top of each other, the resulting interference pattern is 
described by the
cross term 
\begin{eqnarray}
{\cal I}({\bf r},t)&=&C\langle \hat{\psi}({\bf r},
t)e^{-i({\bf k}\cdot {\bf r}%
+\chi )}+\hat{\psi}^{\dagger }({\bf r},t)e^{i({\bf k}\cdot 
{\bf r}+\chi
)}\rangle  \nonumber \\
&=&2C\bar{N}^{1/2}\phi _{\bar{N}}({\bf r})|{\cal F}_{\bar{N}
}(t)|\cos ({\bf k%
}\cdot {\bf r}+\chi -arg({\cal F}_{\bar{N}}(t))).
\end{eqnarray}
In the thermodynamic limit where ${\cal F}_{\bar{N}}
(t)\rightarrow e^{i\eta
} $ this interference pattern takes the form of a 
stationary spatial
modulation of wavevector ${\bf k}$ over the spatial extent 
$r_{\bar{N}}$ of
the condensate (we assume $|{\bf k}|r_{\bar N}>>1$ so that 
an interference
pattern is evident). However, due to the factor
$|{\cal F}_{\bar{N}}(t)|$
appearing in the interference pattern it is clear that the 
interference will
display the collapses and revivals predicted for the 
macroscopic wave
function: when the macroscopic wave function collapses and 
revives so does the visibility of the interference pattern
with the same period $T_{\bar N}$.

Thus quantum interference of two condensates provides an 
experimental
signature of the predicted collapses and revivals. 
Furthermore, the
arguments are easily extended to the case that both 
condensates display
collapses and revivals, that is two identical condensates, 
and the same
conclusions apply.

\subsection{Experimental parameters}

The collapse and revival times calculated for the current 
experiments are
shown in columns six and seven of Table I, respectively. 
Here we see that
except for the weak trap in Rb (second row) the collapse 
times are all less
than one second, indicating that collapse of the 
macroscopic wave function
occurs in these traps. In contrast, the revival times are 
considerably
longer. However, the 6 s revival time found for the strong 
trap in Rb is
still within the condensate lifetime of 15 s quoted for 
that experiment \cite{AndEnsMat95},
though the effects of dissipation may quench the revival. 
In this respect,
we note that the recent Rb experiment (third row)
with a higher trap frequency and mean
particle number has a lower revival time of 3.75 s.

\section{Quantum Phase from Measurements}

\label{section4}

\label{section3} 

In this Section we look at the build up of quantum 
coherence between two
Bose-Einstein condensates initially in Fock states. Since 
we start from Fock
states no phase is initially assumed with a phase only 
established via the
measurement process. The system we consider was first 
proposed by
Javanainen and Yoo\cite{JavYoo96}, and consists of two 
Bose-Einstein
condensates which are dropped on top of one another. Each 
of the condensates
consist of $n$ atoms with momentum
$k_{1}$ and $k_{2}$ 
directed along the x-axis,
respectively. Atoms
are detected on a screen placed below the two condensates. 
A detection at
some position $x$ is represented by the field operator 
for the sum of the two condensates
\begin{equation}
\hat{\Psi}(x)=\frac{1}{\sqrt{2}}
\left(\hat a_{1}+\hat a_{2}e^{i\phi 
\left( x\right)
}\right) ,
\end{equation}
where $\phi (x)=(k_{2}-k_{1})x$, and
$\hat a_{1}$ and $\hat a_{2}$ 
are the atom
annihilation operators for the first and second 
Bose-Einstein condensates
respectively (see below). An interference pattern is 
generated from the two
condensates because every detection of an atom introduces 
uncertainty into
the atom number in each condensate since we do not know 
which condensate
this atom came from. However, the total number in both 
condensates is always
known since it is just the initial total number minus the 
number of
detections we have observed. The uncertainty relationship 
between the
relative atom number and phase between these condensates 
allows us to
establish a relative phase between them to some 
precision.

The effect of collisions on the establishment of quantum 
phase has been
studied previously \cite{Wong}. Here we extend this work to 
demonstrate collapses and revivals in the visibility
pattern between two interfering condensates.

\subsection{Single-mode approximation}

For the purposes of developing a model based on atomic 
detection the full
quantum field theory involving the field operators is 
cumbersome. Thus, here
we reduce the full problem by using a mode expansion and 
truncating. In
particular, we write the Heisenberg field annihilation 
operator as a mode
expansion over single-particle states as 
\begin{eqnarray}
\hat{\psi}({\bf r},t) &=&\sum_{\alpha }\hat{a}_{\alpha }
(t)\varphi _{\alpha
}({\bf r})e^{-i\mu _{\alpha }t/\hbar }  \nonumber \\
&=&\hat{a}(t)\phi _{\bar{N}}({\bf r})e^{-i\mu _{\bar{N}}
t/\hbar }+\tilde{\psi%
}({\bf r},t),
\end{eqnarray}
where $\{\varphi _{\alpha }({\bf r})\}$ are a complete 
orthonormal basis
set, and $\{\mu _{\alpha }\}$ the corresponding 
eigenvalues. Here in the
last line we have chosen $\varphi _{0}({\bf r})=\phi 
_{\bar{N}}({\bf r})$ as
one member of the complete orthonormal set, and identified 
$\mu _{0}=\mu _{%
\bar{N}}$ and $\hat{a}=\hat{a}_{0}$. Then by construction 
the first term in
the mode expansion acts only on the condensate state vector,
whereas the
second term $\tilde{\psi}({\bf r},t)$ accounts for the 
non-condensate atoms.
Then substituting the mode expansion in the 
second-quantized Hamiltonian (%
\ref{QHam}), retaining only the first term representing the 
condensate, and
using the Gross-Pitaevskii Eq. (\ref{GroundEq}),
we obtain the following single-mode quantum Hamiltonian for 
the condensate in the Schr\"odinger picture
\begin{equation}
\hat{H}_S=\hat H(0)=\hat{a}^{\dagger }\hat{a}e_{\bar N}
+\frac{\hbar \kappa }{2}\hat{a}^{\dagger }\hat{a}
^{\dagger }\hat{a}%
\hat{a},
\end{equation}
where
\begin{equation}
\hbar\kappa=U_0\int d{\bf r}|\phi_{\bar N}|^4  ,
\label{kappaDef}
\end{equation}
which using Eq. (\ref{overlap}) yields
$\hbar \kappa=\frac{10}{7}\mu _{\bar{N}}^{\prime }$,
$\kappa$ being the collision rate between
condensate atoms.
The first term on the right-hand-side, which gives
rise to an energy shift, is the number
operator $\hat{a}^{\dagger }\hat{a}$
times the single-particle contribution to
the energy per particle,
\begin{equation}
e_{\bar N}=
\int d{\bf r}\left [\frac{\hbar^2}{2m}
|\nabla\phi_{\bar N}|^2+V({\bf r%
})|\phi_{\bar N}|^2 \right],
\end{equation}
whereas the second term accounts for many-body
interactions.  To proceed we introduce an
interaction picture defined by the transformation
\begin{equation}
\hat H_I(t)=\hat U^\dagger(t)
\left[\hat H_S-i\hbar\frac{\partial\hat U(t)}{\partial t}
\right]\hat U(t)  ,
\qquad
\hat U(t)=e^{-i(e_{\bar N}-\hbar\kappa/2)
\hat{a}^{\dagger }\hat{a}t/\hbar}  .
\end{equation}
which yields the single-mode Hamiltonian
\begin{equation}
\hat{H}(t)
=\frac{\hbar \kappa }{2}(\hat{a}^{\dagger }\hat{a})^2
,  \label{Qhamsing}
\end{equation}
where we have dropped the subscript I on the interaction picture
operators for simplicity in notation.
In the following Sections we shall use this single-mode
Hamiltonian, which is that of a quantum anharmonic
oscillator, as a model for an individual atomic BEC.

A basic assumption underlying this model is that the
quantum state describing the condensate involves only
particle numbers $n\approx\bar N$, that is the particle
number remains close to the mean number.  The single-mode
Hamiltonian (\ref{Qhamsing}) gives rise to a particle
number dependent phase-shift
$\chi_N=\kappa N=\frac{10}{7}\mu_{\bar N}N/\hbar$
(see Sec. V).  However, Eq. (\ref{Foft}) reveals a
particle number dependent phase-shift
$\chi_N=2\mu_{\bar N}N/\hbar$.  The discrepancy
between these two results arises since the phase-shift
in the single-mode theory is an average over the
spatial profile of the macroscopic wave function,
as evidenced by the overlap integral in the
definition of the collision rate $\kappa$ in
Eq. (\ref{kappaDef}).
This averaged value is less than the peak value
of the actual phase-shift which appears in expression
(\ref{Foft}).  In the following we shall
replace the factor $\frac{10}{7}$ by 2 in the
definition of $\kappa$
\begin{equation}
\hbar\kappa=2\mu_{\bar N}  ,
\label{KapDef}
\end{equation}
so that the correct phase-shift is produced by the
quantum anharmonic oscillator model.
This relation provides the connection
between the
single-mode theory parameters and those
obtained from
the treatment based on the Gross-Pitaevskii equation.

In general we shall consider two interfering condensates.
For the $i^{th}$ condensate described by single-mode 
operators $\hat{a}_{i},%
\hat{a}_{i}^{\dagger }$ we assign a wave vector $k_{i}$, 
directed along the
x-axis for simplicity, so that the interference between the 
condensates can
be discussed. The resulting interference pattern would be 
further modulated
by the detailed spatial structure of each condensate, as 
determined by the
solution of the Gross-Pitaevskii equation, but this is of 
secondary
importance and we do not discuss it here. In general, we 
only assume that
any interference is well resolved within the spatial extent 
of the
condensates.

\subsection{Effect of Collisions}

The effect of collisions can be numerically simulated via a 
Monte Carlo wave
function method. The Monte Carlo method is used to simulate 
both the
detection process of rate $\gamma$ (as quantum jumps)
and the time evolutions between
detections (as evolution of the wave function between the 
jumps). The
initial Fock state, say $|n,n\rangle ,$ becomes an 
expansion of $m+1$
entangled Fock states after $m$ detections. Let us write 
the state vector
after $m$ detections as
\begin{equation}
|\varphi _{m}\rangle =\sum_{k=0}^{m}c_{k}|n-m+k,n-k\rangle , 
\label{wfn0}
\end{equation}
which is normalized so that $\sum \left| c_{k}\right| ^{2}
=1$. The effects
of the collisions are included in the time evolution 
operator for the two condensates
\begin{equation}
\hat {\cal U}\left( t\right)=
e^{-i(\hat H_1+\hat H_2)t/\hbar}
=\exp \left( -\frac{i\kappa }{2}
\left[ \left( \hat{a}%
_{1}^{\dag }\hat{a}_{1}\right) ^{2}+\left( \hat{a}_{2}
^{\dag }\hat{a}%
_{2}\right) ^{2}\right] t\right)  ,
\end{equation}
with $\hat H_i$ the Hamiltonian for the $i^{th}$
condensate.
Thus after $m$ detections we have
prepared a state $|\varphi _{m}\rangle$
which then experiences the above evolution
operator for a time $t$ followed by a further detection. 
The conditional
visibility of this detection is 
\begin{equation}
p\left( x|x_{1},\ldots ,x_{m},t\right) \propto \langle 
\varphi _{m}|\hat {\cal U}%
^{\dag }\left( t\right) \hat{\Psi}^{\dag }(x)\hat{\Psi}
(x)\hat {\cal U}\left(
t\right) |\varphi _{m}\rangle .  \label{cond_vis}
\end{equation}
The expectation value for the total number operator is 
\begin{equation}
\langle \varphi _{m}|\hat {\cal U}^{\dag }\left( t\right)
\left(\hat{a}%
_{1}^{\dag }\hat{a}_{1}+\hat{a}_{2}^{\dag }\hat{a}_{2}
\right) \hat {\cal U}\left(
t\right) |\varphi _{m}\rangle =2n-m,  \label{number}
\end{equation}
and, as expected, it is just the total initial
number of atoms minus the number
of detections. The expectation value of the cross term 
which gives the size
of the conditional visibility is 
\begin{eqnarray}
&&\langle \varphi _{m}|\hat {\cal U}^{\dag }\left( t\right) 
\hat{a}_{1}^{\dag }%
\hat{a}_{2}\hat {\cal U}\left( t\right) |\varphi _{m}\rangle 
e^{i\phi \left(
x\right) }  \nonumber \\
&=&\sum_{k=1}^{m}c_{k}^{*}c_{k-1}\sqrt{\left( n-k+1\right) 
\left(
n-m+k\right) }\exp \left\{ i\kappa t\left[ 2k-\left( 
m+1\right) +i\phi
\left( x\right) \right] \right\} .  \label{cross_term}
\end{eqnarray}
Putting Eqs. (\thinspace \ref{number}) and
(\ref{cross_term}) into Eq.
(\thinspace \ref{cond_vis}) we obtain 
\begin{equation}
p\left( x|x_{1},\ldots ,x_{m},t\right) \propto 
n-m/2+\sum_{k=1}^{m}{\cal A}%
\left( k\right) \cos \left[ \phi \left( x\right) +\kappa 
t\left(
2k-m-1\right) -\Theta _{k}\right],  \label{cond_vis2}
\end{equation}
where we have defined the phase $\Theta _{k}$ from the 
coefficients of the
state vector as $c_{k}^{*}c_{k-1}=A_{k}e^{-i\Theta _{k}}$ 
and the weighting
function ${\cal A}$ is 
\begin{equation}
{\cal A}\left( k\right) =A_{k}\sqrt{\left( n-k+1\right) 
\left( n-m+k\right) }%
.  \label{weighting}
\end{equation}
We take $\Theta _{k}$ outside the summation since it is 
defined as the
relative phase between neighboring number states which we 
will assume to be
fairly constant for large $m$. As we detect more and more 
atoms the
resulting entangled state approaches something that 
resembles a coherent
state where the relative phase between neighboring number 
states (in a
number state basis) is identical to the phase of this 
state. Let this fairly
constant relative phase be denoted as $\Theta $ which is a 
good estimate of
the relative phase between the two condensates.
By expanding the cosine terms in Eq.
(\ref{cond_vis2}) to
separate the time dependence from the phase terms,
and noting that the resulting
summation over sine functions vanishes due to the
cancelation of positive and negative
frequency components, we obtain 
\begin{equation}
p\left( x|x_{1},\ldots ,x_{m},t\right) \propto 
n-m/2+\sum_{k=1}^{m}{\cal A}%
\left( k\right) \cos \left[ \left( 2k-m-1\right) \kappa 
t\right] \cos \left[
\phi \left( x\right) -\Theta \right] .  \label{eqn_vis}
\end{equation}
The visibility of the interference pattern is
therefore determined from a weighted
sum over cosines of differing frequencies. These 
frequencies depends on the
parity of the expression enclosed by the round brackets, 
$2k-m-1.$ When this
is odd ($m$ is even) the frequencies are $\left\{ 1-m,
\ldots
,-3,-1,1,3,\ldots ,m-1\right\} $ whereas when it is even 
($m$ is odd) they
are $\left\{ 1-m,\ldots ,-2,0,2,\ldots ,m-1\right\} .$ In 
both cases the
revival period of the visibility is 
\begin{equation}
T=\frac{\pi}{\kappa } = \frac{\pi\hbar}{2\mu_{\bar N}}
,  \label{period}
\end{equation}
with $\bar N=n$.  This revival period is precisely
half that obtained assuming Bose broken symmetry
in Sec. III, and this difference shall be taken up
in Sec. V.
For the case where $m$ is even so that we have a 
sum over odd
frequencies, the cosine term in Eq. (\ref{eqn_vis}) 
alternates between plus
or minus $1$ at subsequent revival times. Since the 
visibility is by
definition a positive quantity, this alternate sign change 
in the cosine
term represents a $\pi $ phase shift of the interference 
pattern at
alternate revivals.

For collision to detection rate ratio of one we expect from 
Eq. (\ref{period}) a period of $\pi$ when time is measured in
units of the inverse detection rate $\gamma$.
This agrees very well to 
the collapse and
revival period displayed in the middle graph of Fig. 
(\ref{fig01}). This
figure and the $\pi $ phase shift will be explained in 
detail in Sec. V.

\subsection{Numerical results}

In a numerical simulation the effects of collisions between 
subsequent
detections can be easily modeled with the use of the Monte 
Carlo wave
function method\cite{MCWF}. We use the following effective 
Hamiltonian 
\begin{equation}
\hat H_{eff}=\frac{\hbar \kappa }{2}
\left[ \left( \hat a_{1}^{\dag }
\hat a_{1}\right)
^{2}+\left( \hat a_{2}^{\dag }\hat a_{2}\right) ^{2}\right] 
-\frac{i\hbar \gamma }{2}%
\left( \hat a_{1}^{\dag }\hat a_{1}
+\hat a_{2}^{\dag }\hat a_{2}\right)  ,
\end{equation}
where the detection and collision rates are
$\gamma $ and $\kappa$ respectively.
The detections are then turned off and the 
system undergoes
coherent evolution by the Hamiltonian 
\begin{equation}
\hat H=\frac{\hbar \kappa }{2}
\left[ \left( \hat a_{1}^{\dag }\hat a_{1}
\right) ^{2}+\left(
\hat a_{2}^{\dag }\hat a_{2}\right) ^{2}\right] .
\label{coll_ham}
\end{equation}
We have modeled the collisions that occur only between the 
atoms of the same
condensate. A cross-collision term ($\hat{a}_{1}^{\dag }
\hat{a}_{1}\hat{a}%
_{2}^{\dag }\hat{a}_{2}$) between the condensates is not 
included. The size
of the coefficient of this term depends on the physical 
geometry of the
situation (the overlap of the two condensates) ranging from 
zero to $\hbar
\kappa $. Setting this coefficient to zero we are
taking the worst case scenario
where the effects of the collisions are the strongest. 
Alternatively, setting the cross-collison term to
$\hbar k$ completes the square in the Hamiltonian
so that the subsequent evolution depends only on the
total atom number. The effect of the collisions would then 
be to rotate the
phase of the entire state vector, and the coherence between 
the individual
entangled number states would be preserved.

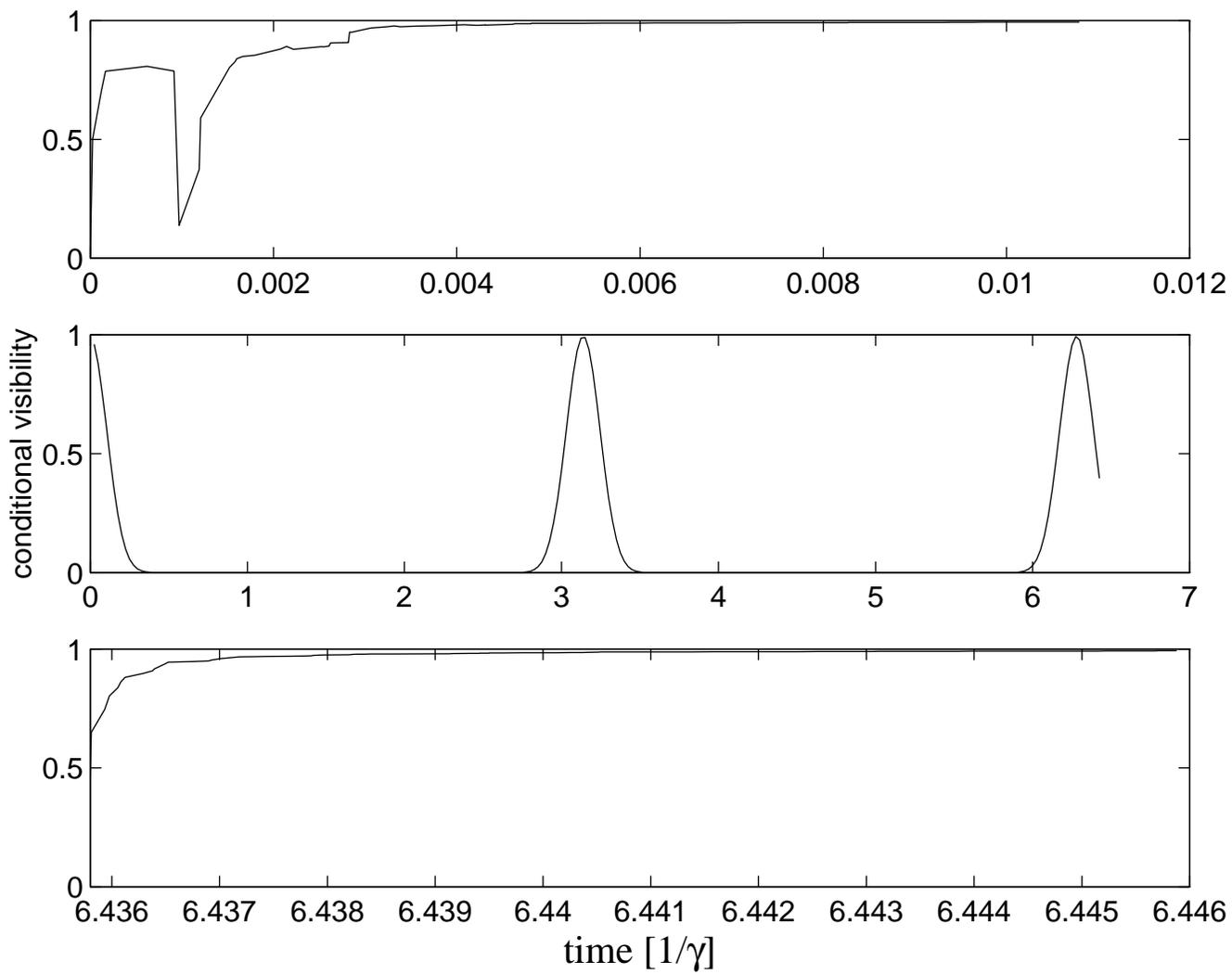
\begin{figure}[p]
\caption{The top graph shows the growth in the conditional 
visibility as a
function of time corresponding to $100$ atomic detections. 
The middle
displays the collapse and revivals of this conditional 
visibility when the
detection process is turned off and the bottom graph shows 
the growth of
this visibility when the detection process is turned on 
again. The total
number of atoms in the two condensates was $10,000$ and we 
have used a
collision to detection rate ratio of one ($\kappa =\gamma 
$) for the initial
and final detection periods.}
\label{fig01}
\end{figure}

In each run of the numerical simulation the state vector 
experiences three
different regimes. Initially a sequence of detections are 
accumulated to
prepare the entangled state, after which the detections
are turned off. During the coherent evolution stage free of
detections, the conditional visibility undergoes 
collapses and
revivals due to collisions. Finally, the detections are 
turned on again. If
the detections are turned on when the visibility is zero, 
during a collapse,
it is quickly re-established by the second sequence of 
detections. If,
however, the detections are turn on when the visibility is 
in a revival phase
the visibility starts from this non-zero value and then 
quickly increases to
one. This behavior is seen in Fig.\thinspace\ref{fig01}
where the detection process occurs at a 
much faster rate than the collapse and revivals, and
we thus graph the single run on three
separate set of axes: The top graph shows the initial 
growth of the
conditional visibility due to the detections of atoms from 
the condensates, and
the visibility quickly increases to a value close to unity
after $100$ atoms are detected.  (In these simulations
the total number of atoms is $10^5$ and 200 atomic
detections are made, so the assumption underlying
the quantum anharmonic oscillator model that the mean
number of atoms varies little is well
obeyed.)
The middle graph shows what happens to the 
visibility once the detection process is turned off and
the state vector undergoes coherent
evolution due to the Hamiltonian in Eq. (\ref{coll_ham}),
and collapses and revivals of the visibility are clearly
evident.  This behavior is reminiscent of the
collapses and revivals in the Jaynes-Cummings model of
quantum optics for a two-photon process which also displays
the periodic revivals of Fig.\thinspace\ref{fig01}.
The bottom graph displays the subsequent evolution of the
visibility when the detection process is turned on again.
Due to the collapses and revivals during the coherent
evolution, the initial visibility for the final stage
depends on the
time at which the detections are reinitiated.
Whatever the initial visibility, this
second detection proceeds very rapidly to increase the 
visibility to a value close to unity.

The cleanliness and exhibition of full revivals in the 
middle graph of Fig.1 can be
understood by noticing that the two condensate system is a 
closed system (no
loss mechanism) undergoing coherent evolution. The state 
vector after the
initial sequence of detections is an expansion of entangled 
number states.
The effect of collisions during the coherent evolution is 
to rotate the
phase of the coefficients of each entangled state by an 
amount proportional
to the sum of the squares of the number of atoms in each 
condensate. Thus
the phase of the coefficients of this state vector rotate 
at differing
frequencies.

So far we have shown collapses and revivals in the 
conditional visibility of
the interference pattern.  For something more relevant to an 
experimental
situation we would like to look at variables associated 
with the actual
observed interference patterns. The phase shift of an 
interference pattern is
a direct measure of the relative phase established between 
the two
condensates. Thus, let us consider the following procedure:
Firstly, we
prepare a state vector of the two condensates with an 
established relative
phase via measurements, and consider this entangled state 
between the two
condensates after the detections to be our initial state 
which possesses
some degree of coherence. This state is then allowed to 
undergo coherent evolution with no detections, and
finally we turn on the measurement
process after an elapsed time and collect our
second sequence of measurements. The
phase of the resulting interference pattern is calculated 
with respect to
the phase of the interference pattern we observed 
previously from the first
sequence of measurements. Now we re-prepare the initial 
state and
repeat the second sequence of measurements and subsequent 
calculation of
phase. Repeating this many times we obtain a set of 
relative phases between
the first and second set of measurements, the idea being 
that if the time elapsed between the two detection
regimes corresponds to some multiple of the full revival time 
then this set of
relative phases should be sharply peaked at zero. For other 
elapsed times we
may expect to see partial revivals. We do not need to 
numerically calculate
the coherent evolution but what we need from the numerical 
simulation is the
coefficients of the prepared state $|\varphi _{m}\rangle .$ 
Its coherent
evolution due to the Hamiltonian $\hat H$ describing the 
collisions previously
given by Eq. (\ref{coll_ham}) is 
\begin{eqnarray}
\left| \varphi _{m}\left( t\right) \right\rangle &=&\exp 
\left( -i\hat Ht/\hbar
\right) \left| \varphi _{m}\right\rangle \nonumber \\
&=&\exp \left( -\frac{i}{2}\kappa \left[ n^{2}+\left( 
n-m\right) ^{2}\right]
t\right) \nonumber \\
&&\times \sum_{k=0}^{m}c_{k}\exp \left( -i\kappa \left[ 
-mk+k^{2}\right]
t\right) \left| n-m+k,n-k\right\rangle .
\end{eqnarray}
We use the phase eigenstates for the atom number difference 
between the two
condensates 
\begin{equation}
\left| \phi \right\rangle =\sum_{n_{1},n_{2}}\exp \left[ 
-\frac{i}{2}\left(
n_{1}-n_{2}\right) \phi \right] \left| n_{1},n_{2}
\right\rangle  ,
\end{equation}
which has a factor of one-half in the exponential since the 
state $|\varphi
_{m}\rangle $ has a fixed total atom number of $2n-m$ so 
that the atom
number difference ($n_{1}-n_{2}$) is quantized in units of 
2. Thus this
factor is required so that $\phi $ is the relative phase 
between the
condensates. The probability distribution of the phase 
$\phi $ after elapsed time $t$ is 
\begin{equation}
\left| \left\langle \phi \right| \varphi _{m}\left( 
t\right) \rangle \right|
^{2}=\left| \sum_{k=0}^{m}c_{k}\exp \left( -i\left[ k\left( 
k-m\right)
\kappa t+2k\phi \right] \right) \right| ^{2}.
\end{equation}
This probability is a function of two variables, phase and 
elapsed time, and is
evaluated numerically. We display the probability 
distributions in
``birds-eye'' plots via the ``image'' command using 
MATLAB$^{\copyright }$.
Figure (\ref{fig02}a) displays the probability distribution 
when we have
made an even number of detections ($m$ even). The white 
regions denotes the
peaks with the black ones corresponding to the valleys. We 
see a sharp peak
at zero time about zero phase difference between the two 
interference
patterns, this is not surprising since no coherent evolution 
has occurred. By
the time we have evolved for one revival period we obtain a 
sharp peak at $%
\pi $ radians corresponding to the first revival time with 
$\pi $ phase
shift due to detecting an even number of atoms described in 
the previous
section. The second revival occurs at a phase difference of 
zero radians as
predicted. Away from these revival times, the phase 
distribution is not flat
but displays many partial revivals. Figure (\ref{fig02}b) 
is a zoomed in
view of Fig. (\ref{fig02}a) between $0.2$ and $0.5$ revival 
periods, we can
clearly see the partial revival at $0.2$ which consists of 
five peaks. In
fact we see partial revivals at every integer fraction of a 
revival period
provided the resolution is good enough. We illustrate this 
by labelling to
the left of the graph with the appropriate integer 
fractions corresponding
to the particular partial revival. Figure (\ref{fig03}) 
shows the
distribution for the other case where the number of 
detections is odd. As
predicted there is no $\pi $ phase shifts. Again, we see 
partial revivals
when we zoom in between $0.2$ and $0.5$ revival periods as 
shown in Fig. (%
\ref{fig03}b).

\section{Quantum oscillator model with broken symmetry}

We will show in this brief Section that collapses and 
revivals also arise for the quantum anharmonic
oscillator model
for initial coherent states instead of 
preparing an
entangled state from an initial detection process. In the 
language of the
previous Section we are considering the interference 
between two coherent
states including the effects of collisions. Thus we treat 
the Bose-Einstein
condensates as coherent states, analogous to the treatment
in Sec. III, imposing on them a relative 
phase whereas
previously we establish this phase via measurements.

The Hamiltonian for the two condensates is
\begin{equation}
H=\frac{1}{2}\sum_{i=1}^{2}\hbar \kappa \left( \hat{a}_{i}
^{\dag }\hat{a}%
_{i}\right) ^{2}  ,
\end{equation}
giving the
Heisenberg equation of motion for each field anihillation
operator
\begin{equation}
\frac{d\hat{a}_{i}}{dt}=\frac{1}{i\hbar }\left[ \hat{a}
_{i},H\right] \\
=\frac{\kappa }{2i}\left( 1+2\hat{a}_{i}^{\dag }\hat{a}
_{i}\right) \hat{a}%
_{i}  .
\end{equation}
By inspection the time dependence of the $\hat{a}_{i}$ 
operator is 
\begin{equation}
\hat{a}_{i}\left(t\right) =\exp \left[-\frac{i}{2}\left( 
1+2\hat{a}_{i}^{\dag }\hat{a}_{i}\right) \kappa t\right]
\hat{a}_{i}  ,
\end{equation}
which yields the Heisenberg picture field operator 
for the sum of the two modes
\begin{equation}
\hat{\Psi}\left( t\right) =\frac{1}{\sqrt{2}}\left\{ \exp 
\left[ -\frac{i}{2}\left( 1+2\hat{a}_{1}^{\dag }
\hat{a}_{1}\right) \kappa 
t\right] \hat{a}_{1}+\exp \left[ -\frac{i}{2}
\left( 1+2\hat{a}_{2}^{\dag }
\hat{a}_{2}\right)
\kappa t\right] \hat{a}_{2}\right\}  ,
\end{equation}
where we have suppressed the spatial dependence.
This yields the operator for the
intensity of the atomic pattern as
\begin{equation}
\hat{\Psi}^{\dag }\left( t\right) \hat{\Psi}\left( t\right) 
=\frac{1}{2}%
\left\{ \hat{a}_{1}^{\dag }\hat{a}_{1}+\hat{a}_{2}^{\dag }
\hat{a}_{2}+\hat{a}%
_{1}^{\dag }\exp \left[ i\left( \hat{a}_{1}^{\dag }\hat{a}
_{1}-\hat{a}%
_{2}^{\dag }\hat{a}_{2}\right) \kappa t\right] \hat{a}_{2}
+\text{h.c.}%
\right\} .  \label{NumDiff}
\end{equation}
Thus, if we define
the initial coherent states as $|\alpha \rangle $ 
and $|\beta\rangle$
for the modes $\hat{a}_{1}$and $\hat{a}_{2}$ 
respectively, the
intensity is evaluated to be 
\begin{eqnarray}
I &\propto &\left\langle \alpha ,\beta \right| \hat{\Psi}
^{\dag }\left(
t\right) \hat{\Psi}\left( t\right) \left| \alpha ,\beta 
\right\rangle \nonumber \\
&=&\frac{1}{2}\left\{ \left| \alpha \right| ^{2}+\left| 
\beta \right|
^{2}+\alpha ^{*}\beta \exp \left[ \left( e^{i\kappa t}
-1\right) \left|
\alpha \right| ^{2}+\left( e^{-i\kappa t}-1\right) \left| 
\beta \right|
^{2}\right] \right. \nonumber \\
&&\left. +\beta ^{*}\alpha \exp \left[ \left( e^{-i\kappa t}
-1\right) \left|
\alpha \right| ^{2}+\left( e^{i\kappa t}-1\right) \left| 
\beta \right|
^{2}\right] \right\}  ,  \label{term0}
\end{eqnarray}
with the aid of the following identity \cite{Louisell} 
\[
\left\langle \alpha \left| e^{-xa^{\dag }a}\right| \alpha 
\right\rangle
=\exp \left[ \left( e^{-x}-1\right) \left| \alpha \right| 
^{2}\right] . 
\]
In the case of maximum visibility of the interference 
pattern we need the
modes to be equal in amplitude, thus we set $\beta =\alpha 
e^{-i\phi }$
where $\phi $ is the phase difference between the modes.
Substituting this
into Eq.\thinspace (\ref{term0}) we obtain 
\begin{equation}
I\propto \left| \alpha \right| ^{2}\left\{ 1+\exp \left[ 
2\left| \alpha
\right| ^{2}\left( \cos \kappa t-1\right) \right] \cos \phi 
\right\} .
\label{coll_rev}
\end{equation}
The characteristics of collapses and revivals can be 
clearly seen in
Eq.\thinspace (\ref{coll_rev}). Revivals occur at times 
$t=2\pi n/\kappa $
(where $n$ is an integer) with the shape of the collapses 
described by the
exponential term whenever $t$ is no longer a multiple of 
$2\pi /\kappa $.
The period of these revivals is 
\begin{equation}
T=\frac{2\pi }{\kappa}=\frac{\pi\hbar}{\mu_{\bar N}},  \label{CohPeriod}
\end{equation}
which is identical to Eq. (\ref{RevPer}) with $\bar N=|\alpha|^2$.
This gives twice the period of the previous case of 
revivals from a
detection process (see Eq. (\ref{period})), even though we 
have used
identical Hamiltonians. This difference arises because we 
have two
independent coherent states here whereas in the case of 
detecting atoms from
two initial Fock states the total atom number is always 
fixed at the initial
total number minus the number of atomic detections. This 
additional
constraint on the total number means that the difference 
atom number
operator inside the exponential in Eq. (\ref{NumDiff}) is 
quantized in units
of $2.$ The above expression for the period is only 
applicable when we are
free to superpose total atom numbers, as in the case of two 
coherent states
for which the difference atom number operator is quantized 
in units of $1.$
This gives a factor of $2$ difference in the revival times 
between assuming
an initial relative phase and establishing this phase in 
the dynamics of the
revivals. The visibility, as described by Eq. 
(\ref{coll_rev}), smoothly
drops to its minimal value halfway between subsequent 
revivals, exactly
where the state induced by detection would have an 
additional revival.

\section{Summary and conclusions}

In this paper we have shown that within the approximation
of Bose broken symmetry the macroscopic wave 
function in small atomic samples exhibits collapse and
revivals in time while the BEC is maintained in the form
of ODLRO. For current experiments the collapse time is a
second or less.  The revival time is longer but our
results show that it may still fall within the
condensate lifetime for some experiments.
To detect the collapses and revivals
experimentally a scheme is required which is sensitive to 
the macroscopic wave function directly, and this does not
seem to be the case for the coherent light scattering
methods previously discussed \cite{Jav94}-\cite{JavRuo95}.
However, Imamo\={g}lu and Kennedy 
\cite{ImaKen96} and Javanainen \cite{Jav96} have recently
proposed light scattering schemes involving two
independent condensates coupled by a common excited state. 
These schemes rely on the fact that when one condensate is driven 
optically the light scattered from the other condensate has a
non-zero value of the electric
field and a phase proportional to the relative phase of the 
two condensates.  By driving both wells and adjusting the phase
difference of the fields the scattering can be suppressed via
quantum interference, and this in turn determines the phase
difference between the two condensates. The scattering
is therefore sensitive to Bose broken gauge symmetry. In 
addition, the light scattering rate is proportional to the
magnitude of the macroscopic wave function, so these schemes
could be used to detect the collapses and revivals experimentally.
Alternatively, as we have shown here, a direct measurement
of the collapse and revivals in the visibilty of the interference
between two condensates can be made.

In a second approach we have also studied the establishment
of a relative phase between two interfering condensates using
the explicit measurement model first proposed by
Javanainen and Yoo \cite{JavYoo96}.  This model is free
from any assumptions concerning Bose broken symmetry or
the thermodynamic limit, and is therefore applicable
to small atomic samples.  In this case we consider the
two interfering condensates to be in number states
initially, an extreme example for which there is no
relative phase before atomic detections.
These condensates are then prepared into an 
entangled state vector composed of entangled number
eigenstates via the measurement process. The
effects of collisions in the time evolution is to rotate 
the phases of the individual entangled eigenstates of the
state vector. We observe the
collapse and revivals of this state under coherent 
evolution when the
measurement process is turned off. Accurate predictions of 
the period of
these collapses and revivals were obtained. The simple 
anharmonic model of
interference between two condensates also displays 
collapses and revivals.
The period of these revivals is twice the time required for 
the previous
case since we have neglected the constraint of the total 
atom number being
fixed.

\section*{Acknowledgements}

Dan Walls acknowledges support from the Office 
of Naval
Research, the New Zealand Foundation for Research Science 
and Technology, and the Marsden Fund
Useful discussions with S. Barnett, K.
Burnett, and J. C. Garrison are appreciated.
\newpage

\begin{center}
\begin{tabular}{|*{8}{c|}} \hline
Expt. & $\omega_\perp^0$ & $\lambda$ & $a$ & $\bar N$ & 
$\zeta$ & $t_{coll}$ & 
$T_{\bar N}$ \\ \hline\hline
1 & 471 rad s$^{-1}$ & 2.8 & 5.2 nm & 2000 & 0.1 & 0.13 s & 
6 s \\ \hline
2 & 51 rad s$^{-1}$ & 2.8 & 5.2 nm & 2000 & 0.15 & 2 s & 86 
s \\ \hline
3 & 828 rad s$^{-1}$ & 2.8 & 5.2 nm & 4500 & 0.05 & 55 ms & 
3.75 s \\ \hline
4 & 2010 rad s$^{-1}$ & 0.056 & 4.9 nm & $5\times 10^6$ & 
0.02 & 0.32 s & 710
s \\ \hline
\end{tabular}
\end{center}

\vspace{0.25cm}

\noindent 1. "Strong trap" experiment in Rb from Ref. 
\cite{AndEnsMat95} 
\newline
2. "Weak trap" experiment in Rb from Ref. \cite{AndEnsMat95}
 \newline
3. Experiment in Rb from Ref. \cite{JinEnsMat96} \newline
4. Experiment in Na from Ref. \cite{MewAndDru96b} 

\begin{figure}[p]
\caption{Plot of the phase distribution when the number of 
detections ($m$)
is even as a function of the turning on time. The phase is 
the relative
phase between the first and second sequence of measurements 
in units of
radians while the turning on time is in units of the 
revival period. The
brightness of a region corresponds to the relative 
probability of obtaining
a particular phase for a particular turning on time. The 
bright regions
denotes peaks while the darker ones correspond to valleys. 
The entire plot
is shown in (a) with a zoomed in plot between $0.2$ to 
$0.5$ times displayed
in (b).}
\label{fig02}
\end{figure}

\begin{figure}[p]
\caption{Plot of the phase distribution when the number of 
detections ($m$)
is odd as a function of the turning on time. The entire 
plot is shown in (a)
with a zoomed in plot between $0.2$ to $0.5$ times 
displayed in (b).}
\label{fig03}
\end{figure}

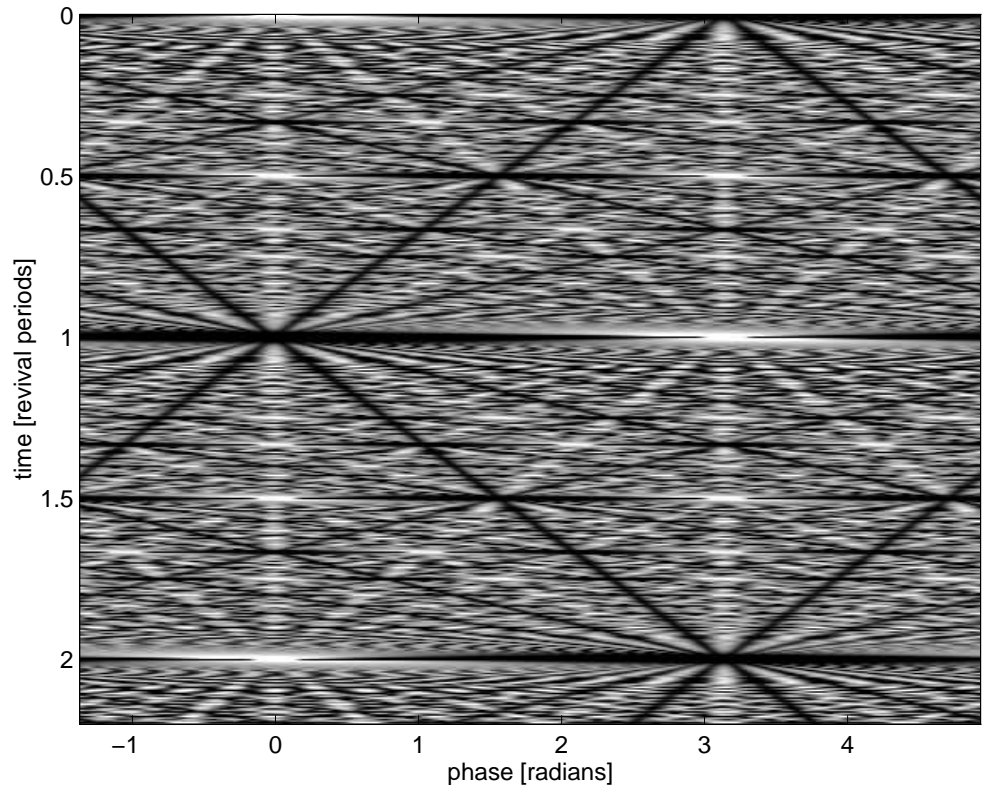
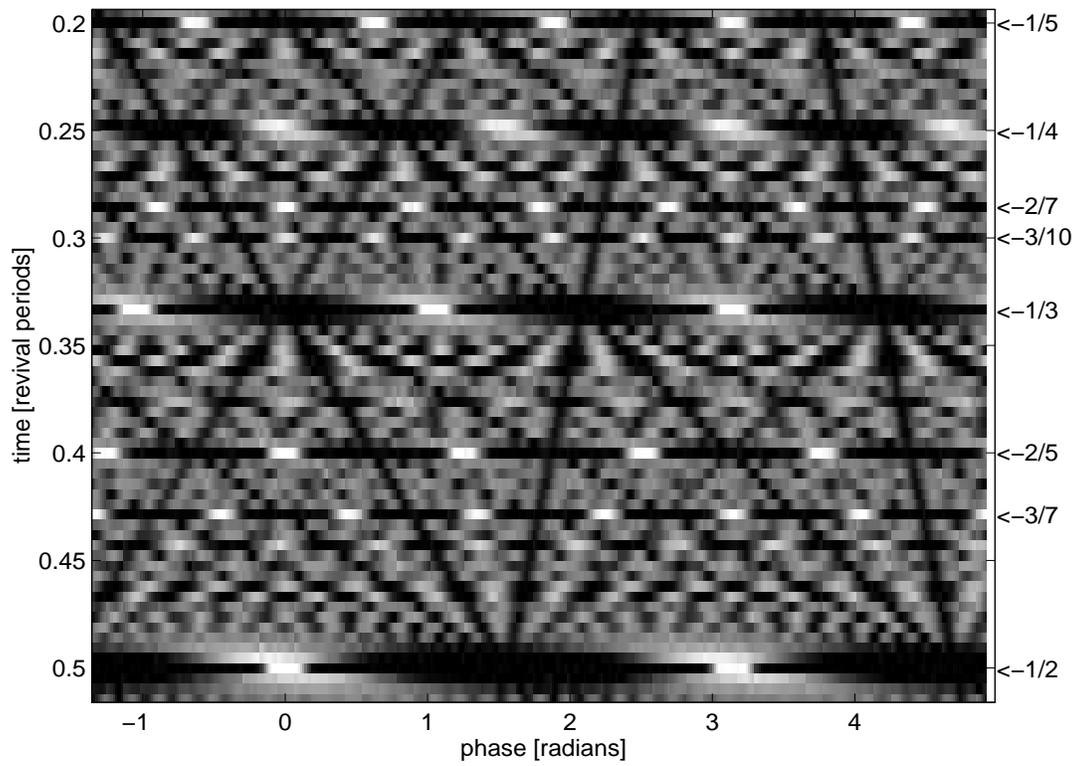
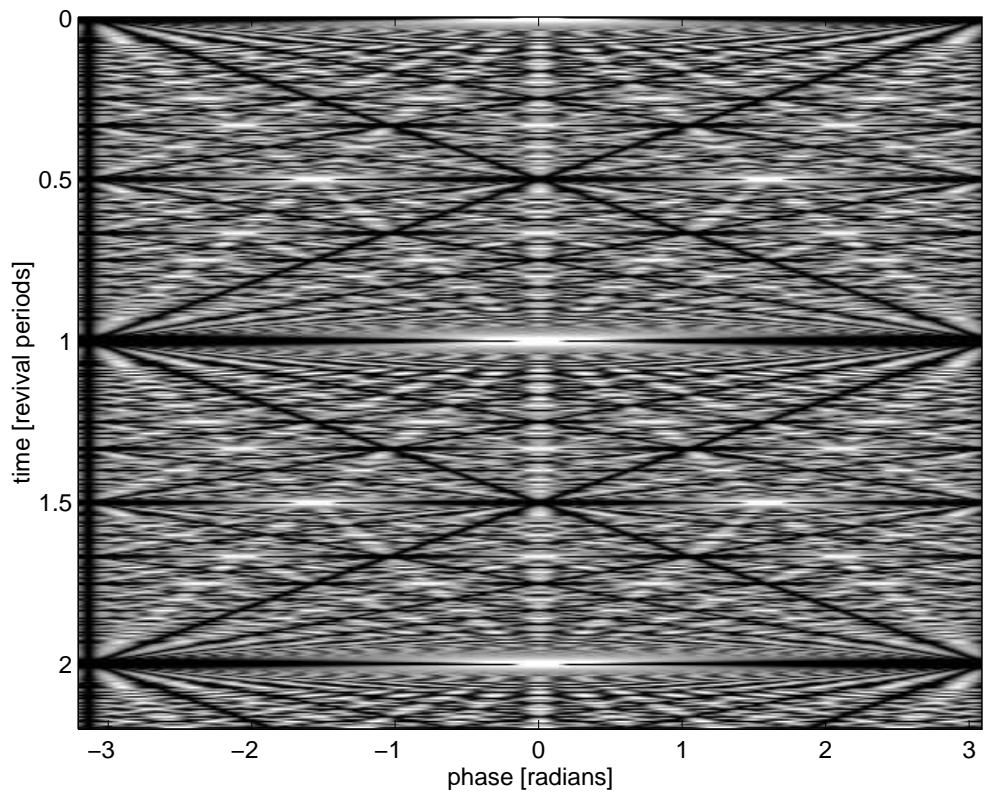
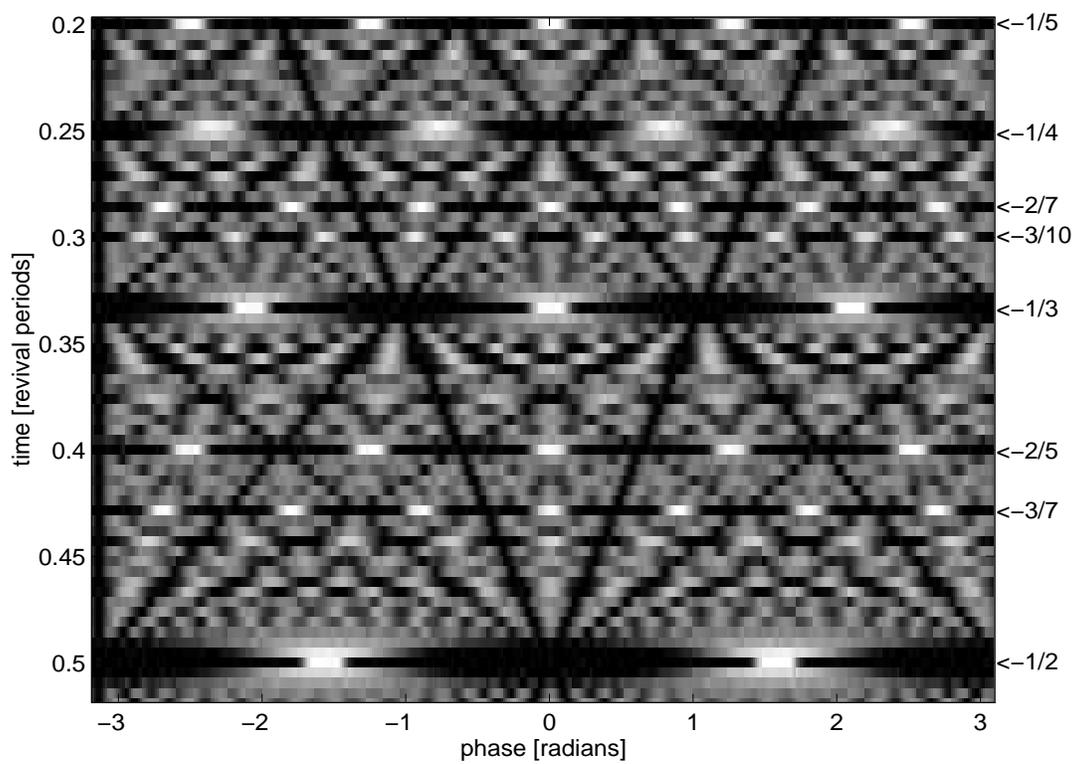
\end{document}